\documentclass{PoS}
\usepackage{amsmath}
\usepackage{amsfonts}

\title{Calculation of the ratios and absolute rates of the $\Xi_b^- \to \pi^- (D_s^- ) \ \Xi_c^0 (2790) \left(\Xi_c^0 (2815) \right)$ and $\Xi_b^- \to \bar{\nu}_l l \ \Xi_c^0 (2790) \left(\Xi_c^0 (2815) \right)$ decays}

\ShortTitle{Calculation of the ratios and absolute rates of $\Xi_b$ decay to $\Xi_c^{(*)}$}

\author{\speaker{R. Pavao},$ˆa$ W-H Liang,$ˆb$ J. Nieves,$ˆc$ E. Oset$a$\\
\llap{$ˆa$} Departamento de
F\'{\i}sica Te\'orica and IFIC, Centro Mixto Universidad de
Valencia-CSIC Institutos de Investigaci\'on de Paterna, Aptdo.
22085, 46071 Valencia, Spain\\
\llap{$ˆb$} Department of Physics, Guangxi Normal University,
Guilin 541004, China\\
\llap{$ˆc$} Instituto de F\'\i sica Corpuscular (IFIC), Centro Mixto
CSIC-Universidad de Valencia, Institutos de Investigaci\'on de
Paterna, Apartado 22085, E-46071 Valencia, Spain \\ 
E-mail: \email{rpavao@ific.uv.es}, \email{liangwh@gxnu.edu.cn},
\email{jmnieves@ific.uv.es}, \email{oset@ific.uv.es}}

\abstract{In this work we calculate the ratios of rates of the $\Xi_b$ nonleptonic and semileptonic decays into the $\Xi_c$(2790) and $\Xi_c$(2815) ($\Xi_c^*$) resonances. These resonances are dynamically generated from the pseudoscalar-baryon and vector-baryon interactions, whose mixing is done using the chiral Weinberg-Tomozawa (WT) meson-baryon interaction extended to four flavors. The first part of the decay is a weak decay that we analyze through their quark constituents where it is noted that only the heavy quarks ($b$ and $c$) participate in the interaction, leaving the light pair ($ds$) as spectators. This first decay then produces a meson-baryon pair that creates the $\Xi_c^*$ through the WT interaction. We then proceed to calculate the decay rates to $\Xi_c$(2790) and $\Xi_c$(2815) for both the nonleptonic and semileptonic cases and then calculate the ratios between them. We do this calculation nonrelativistically and fully relativistically and notice that, even though both approaches yield somewhat different results in the rates, the ratios are very similar (difference on the order of 1\%) in both cases. The absolute values of the decay rates are also successfully calculated by obtaining the rates between our decays and $\Lambda_b \to \pi (D_s ) \ \Lambda_c (2595) \left(\Lambda_c (2625) \right)$ and $\Lambda_b \to \bar{\nu}_l l \ \Lambda_c (2595) \left(\Lambda_c (2625) \right)$ for which there are experimental results, since the momentum transfer is similar such we can cancel out the influence of the quark wave functions in the ratios.}

\FullConference{XVII International Conference on Hadron Spectroscopy and Structure - Hadron2017\\
		25-29 September, 2017\\
		University of Salamanca, Salamanca, Spain}

\begin{document}

\section{Introduction}

The most renowned resonance explained using the unitary coupled channels formalism in chiral perturbation theory is one of the two $\Lambda(1405)$ states \cite{ollerulf,jido,ulfg2,carmen}. It was found to couple mostly to $\bar{K}N$. Similarly the $\Lambda_c(2595)$ has been considered an analog to the $\Lambda(1405)$ \cite{lutz,mizutani}, and it was thought to couple mostly to the $DN$. However, in \cite{weihong,weisemi}, by studying the decays  $\Lambda_b \to \pi (D_s ) \ \Lambda_c (2595) \left(\Lambda_c (2625) \right)$ and $\Lambda_b \to \bar{\nu}_l l \ \Lambda_c (2595) \left(\Lambda_c (2625) \right)$ mixing pseudoscalar-baryon (PB) and vector-baryon (VB) channels, it was discovered that the $D^* N$ channel also plays a very important role in the generation of $\Lambda_c (2595)$. There, the ratio between the partial withs of $\Lambda_c (2595)$ and $\Lambda_c (2625)$ were calculated and found in good agreement with experiment, while without the VB channel the results were in total disagreement.

In this work we extend the ideas of \cite{weihong,weisemi} to the study of the $\Xi_b^- \rightarrow \pi^- \ \Xi_c^0 (2790) (\frac{1}{2}^-)$, $\Xi_b^- \rightarrow \pi^- \ \Xi_c^0 (2815) (\frac{3}{2}^-)$, $\Xi_b^- \rightarrow D_s^- \Xi_c^0 (2790)$, $\Xi_b^- \rightarrow D_s^- \Xi_c^0 (2815)$, $\Xi_b^- \rightarrow \bar{\nu}_l l \ \Xi_c^0 (2790)$ and  $\Xi_b^- \rightarrow \bar{\nu}_l l \ \Xi_c^0 (2815)$ decays. For the PB and VB mixing we use the formalism created in \cite{romanets}, where the Weinberg-Tomozawa interaction was extended to four flavors in a way compatible with heavy quark spin symmetry. We apply these methods to the full hadronic decays and then extend it to the semileptonic decays.

\section{Formalism}
 
\subsection{Hadronic decay}

\begin{figure}[h!]
  \centering
  \includegraphics[scale = 0.58]{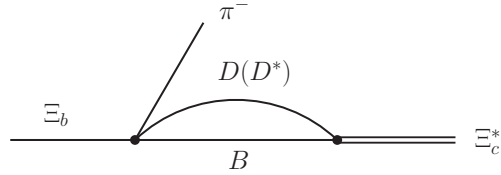}
  \caption{Mechanism for the production of the $\Xi_c^*$ resonances by re-scattering of  $D \left(D^* \right) \Sigma \left(\Lambda\right)$ and coupling of the meson-baryon components to  $\Xi_c^*$.}
  \label{fig:decay3}
\end{figure}
\begin{figure}[h!]
\centering
  \includegraphics[scale = 0.58]{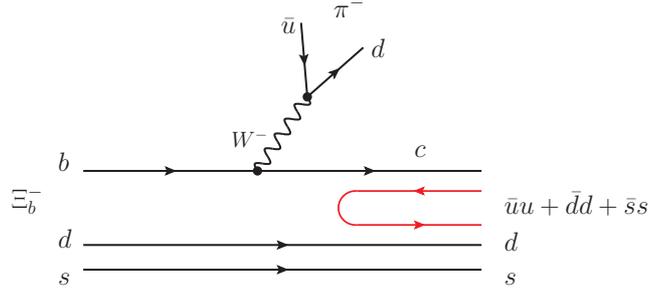}
  \caption{Diagrammatic representation of the weak decay and hadronization after the weak process to produce a meson-baryon pair in the final state.}
  \label{fig:decay2}
\end{figure}

The complete decay mechanism for the reaction $\Xi_b^- \to \pi^- \ \Xi_c^*$ is shown in Fig. \ref{fig:decay3}, where a pion is emitted through weak interaction and then a meson-baryon pair is produced through hadronization. Then the pair re-interacts to form the $\Xi_c^*$ resonance. The pion emission vertex corresponds to the weak process in Fig. \ref{fig:decay2}, where the light quarks $ds$ are only spectators and do not participate in the weak process.

The pion emission vertex was studied in Ref. \cite{weihong} and, at the macroscopical level, its amplitude is
\begin{equation}
\label{Eq:13}
V_P \sim \left\{ \left(i \frac{q^0}{q} \vec{\sigma} \cdot \vec{q} + i q \right) \delta_{J,\frac{1}{2}}-i \frac{q^0}{q} \sqrt{3} \ \vec{S}^+ \cdot \vec{q} \ \delta_{J,\frac{3}{2}} \right\} \text{ME}(q),
\end{equation}
where $\vec{S}^+$ is the spin transition operator from spin $\frac{1}{2}$ to spin $\frac{3}{2}$, such that
\begin{equation}
\left < M' \right| S^+_{\mu} \left| M \right> = \mathcal{C}(\frac{1}{2}, \, 1, \, \frac{3}{2};\, M, \, \mu, \, M'),
\end{equation}
where $\mathcal{C}(\frac{1}{2}, \, 1, \, \frac{3}{2};\, M, \, \mu, \, M')$ are the Clebsch-Gordan coefficients. The ME$(q)$ is the quark matrix element involving the radial wave functions,
\begin{equation}
\label{Eq.:matel}
\text{ME}(q)=\int {\rm d}r \ r^2 j_1(qr) \phi_{\text{in}}(r) \phi^*_{\text{fin}}(r),
\end{equation}
where $j_1(qr)$ is a spherical Bessel function and $\phi_{\text{in}}(r)$ is the radial wave function of the $b$ quark and $\phi_{\text{fin}}(r)$ the radial wave function of the $c$ quark, which is in an excited $L=1$ state. This is because, since the final $\Xi_c^*$ will have negative parity, and the $ds$ pair has positive parity, then the negative parity must come from the $c$ quark (prior to the hadronization) and hence it must have $L=1$.

Next is the hadronization that proceeds as shown in Fig. \ref{fig:decay2}, where a $\bar{q}q$ state is created from the vacuum. Then, we can reorganize the quark states in flavor space and write them in terms of both the physical mesons and the mixed antisymmetric representations of the $\Sigma^-$,  $\Sigma^0$, $\Lambda^0$ baryon states \cite{close}. Here one needs to be careful to write the states using a phase convention coherent with the chiral Lagrangians.
Then, after the hadronization we have, in the isospin basis,
\begin{equation}
\label{Eq.:isoH}
\left|H' \right\rangle = -\sqrt{\frac{3}{2}} \left| \Sigma D (I=\frac{1}{2}) \right\rangle+\frac{1}{\sqrt{6}} \left| \Lambda D (I=\frac{1}{2}) \right\rangle.
\end{equation}

Now we need to study the spin structure of the hadronization. The $\bar{q}q$ pair created from the vacuum has $J^P=0^+$, which means that they need to have $L=1$ and $S=1$. Also the $c$ quark before the hadronization will have the same total angular momentum as the $\Xi_c^*$. 
In the hadronization the total angular momentum of the $c$ and $\bar{q}$ recombine to give the total angular momentum ($j=0,1$) of the meson $D$ or $D^*$. Since the spin of the $ds$ pair is zero, the spin of $\Xi_c^*$ is determined by the total angular momentum of $q$. This recombination was done in Ref.\cite{weihong} using Racah coefficients, such that
\begin{equation}\label{Eq.:CGcoef}
\left|J M; c \right> \left|0 0; \bar{q}q \right>_{^3P_0} \left|0  0; ds \right> 
  = \sum_j \mathcal{C}(j,J) \left|J, \, M; \text{meson-baryon}\right>,
\end{equation}
where the coefficients $\mathcal{C}(j,J)$ are given in Table \ref{tab:tab1}.
\begin{table}[tbp!]
\centering
\caption{$\mathcal{C}(j,J)$ coefficients in Eq. \eqref{Eq.:CGcoef}.\label{tab:tab1}}
\begin{tabular}{l|cc}
\hline\hline
$\mathcal{C}(j,J)$ & $J=\frac{1}{2}$ & $J=\frac{3}{2}$ \\ \hline
(pseudoscalar) $j=0~$ & $\frac{1}{4 \pi} \frac{1}{2}$ & 0 \\
(vector) $j=1$ & ~~$\frac{1}{4 \pi} \frac{1}{2\sqrt{3}}$~~ & $-\frac{1}{4 \pi} \frac{1}{\sqrt{3}}$ \\
\hline\hline
\end{tabular}
\end{table}
Combining Eqs. \eqref{Eq.:isoH} and \eqref{Eq.:CGcoef} we get the amplitude of the hadronization and the resonance formation:
\begin{eqnarray}\label{Eq.:sum12}
   J=\frac{1}{2}: \ V_{\text{had}}(J) &=&  \frac{1}{2} \left(-\sqrt{\frac{3}{2}}\right) g_{R,\Sigma D}\,G_{\Sigma D}  + \frac{1}{2} \frac{1}{\sqrt{6}}  g_{R,\Lambda  D}\,G_{\Lambda D} \nonumber \\
   &&  + \frac{1}{2\sqrt{3}}\left(-\sqrt{\frac{3}{2}}\right) g_{R,\Sigma D^*}\,G_{\Sigma D^*} + \frac{1}{2\sqrt{3}}  \frac{1}{\sqrt{6}} g_{R,\Lambda  D^*}\, G_{\Lambda D^*},
\end{eqnarray}
and
\begin{equation}
\label{Eq.:sum32}
J=\frac{3}{2}: \ V_{\text{had}}(J) = \frac{1}{\sqrt{3}} \left(-\sqrt{\frac{3}{2}}\right)g_{R,\Sigma D^*}\, G_{\Sigma D^*} + \frac{1}{\sqrt{3}}  \frac{1}{\sqrt{6}} g_{R,\Lambda  D^*}\,G_{\Lambda D^*},
\end{equation}

where $G_{BD}$, $G_{BD^*}$ are the loop functions for the propagator of $BD \left(B D^*\right)$, and $g_{R,BD \left(BD^*\right)}$ the coupling of these states to the resonance $\Xi_c^*$ that are calculated in Ref. \cite{romanets}. 

The width for the $\Xi_b \rightarrow \pi^- \Xi^*_c$ decay is given by
\begin{equation}
\Gamma_{\Xi_b \rightarrow \pi^- \Xi_c^*} = \frac{1}{2 \pi}\, \frac{M_{\Xi_c^*}}{M_{\Xi_b}} \;q \; \overline{\sum} \sum |t|^2=
\end{equation}
\begin{equation}
\label{eq:pionwidth}
= \frac{C^2}{2 \pi}\, \frac{M_{\Xi_c^*}}{M_{\Xi_b}} \;q \; \left[\left(q^2+\omega^2_{\pi} \right)\delta_{J,\frac{1}{2}} + 2\omega^2_{\pi} \delta_{J,\frac{3}{2}}\right] |V_{\text{had}}(J)|^2,
\end{equation}
where $C$ contains the matrix element ME$(q)$ and the weak interaction constants and can be assumed to be constant. The $\omega_{\pi}$ is the pion energy.

The case of $D_s^-$ production is identical, with the only difference being that the momentum of the $D_s^-$ is smaller than in the case of pion production. 

\subsection{Semileptonic decay}
The extension to the semileptonic cases $\Xi_b \rightarrow \bar{\nu}_l l \Xi_c^0(2790)$ and $\Xi_b \rightarrow \bar{\nu}_l l \Xi_c^0(2815)$ is straightforward since the main difference is that instead of a $\pi^-$ we have $\bar{\nu}_l l$ production. Combining the $W \bar{\nu}_l l$ and $W c b$ vertices we get
\begin{equation}
t' \propto L^{\alpha} Q_{\alpha},
\end{equation}
with,
\begin{subequations}
\begin{align}
&  L^{\alpha} = \bar{u}_l \gamma^{\alpha} (1-\gamma_5) u_{\nu_l}, \\
&  Q_{\alpha} = \bar{u}_c \gamma_{\alpha} (1-\gamma_5) u_{b},
\end{align}
\end{subequations}
The mass distribution is given by
\begin{equation}
\label{Eq.:ratiomass}
\frac{{\rm d} \Gamma}{{\rm d}M_{\text{inv}}(\bar{\nu}_l l)}= \frac{M_{\Xi_c^*}}{M_{\Xi_b}} 2 m_{\nu} 2 m_l \frac{1}{\left(2 \pi \right)^3} p_{\Xi_c^*} \tilde{p}_l \overline{\sum}\sum |t'|^2,
\end{equation}
where $p_{\Xi_c^*}$ is the $\Xi_c^*$ momentum in the $\Xi_b$ rest frame and $\tilde{p}_l$ the lepton momentum in the $\bar\nu_l l$ rest frame, and $\overline{\sum}\sum |t'|^2$ is given by \cite{weisemi}
\begin{equation}
\overline{\sum}\sum |t'|^2 = C'^2 \frac{8}{m_{\nu} m_l} \frac{1}{M ^2_{\Xi_b}} \left(\frac{M_{\text{inv}}}{2}\right)^2 \left[\tilde{E}^2_{\Xi_b}-\frac{1}{3} \tilde{\vec{p}\,}^2_ {\Xi_b}\right] |V_{\text{had}}(J)|^2,
\end{equation}
where $C'$ is again a factor that contains the matrix element ME$(q)$.

The magnitudes $\tilde{E}_{\Xi_b}$ and $\tilde{\vec{p}}_{\Xi_b}$ are evaluated in the rest frame of the $\bar\nu_l l$ pair and are given by
\begin{subequations}
\begin{align}
& \tilde{E}_{\Xi_b} = \frac{M^2_{\Xi_b}+M^2_{\text{inv}}-M^2_{\Xi_c^*}}{2 M_{\text{inv}}}, \\
& \tilde{p}_{\Xi_b} = \frac{\lambda^{\frac{1}{2}}\left(M^2_{\Xi_b},M^2_{\text{inv}},M^2_{\Xi_c^*} \right)}{2 M_{\text{inv}}}.
\end{align}
\end{subequations}

\section{Results}
We proceed now to show the ratios of rates obtained in \cite{pavao}. 
For the nonleptonic emission, using Eq.\eqref{eq:pionwidth} we get,
\begin{equation}
\label{Eq.:ratiorate}
\frac{\Gamma_{\Xi_b \rightarrow \pi^- \Xi_c (1)}}{\Gamma_{\Xi_b \rightarrow \pi^- \Xi_c (2)}}= \frac{M_{\Xi_c (1)} \;p_{\pi}(1) \;\overline{\sum}\sum |t|^2(1)}{M_{\Xi_c (2)} \;p_{\pi}(2)\; \overline{\sum}\sum |t|^2(2)} = 0.384,
\end{equation}
\begin{equation}
\label{Eq.:ratiorate2}
\frac{\Gamma_{\Xi_b \rightarrow D_s^- \Xi_c (1)}}{\Gamma_{\Xi_b \rightarrow D_s^- \Xi_c (2)}}= \frac{M_{\Xi_c (1)} \,p_{D_s^-}(1) \,\overline{\sum}\sum |t|^2(1)}{M_{\Xi_c (2)} \,p_{D_s^-}(2) \,\overline{\sum}\sum |t|^2(2)} =0.273,
\end{equation}
and, if we assume that $\pi^-$ and $D_s^-$ have similar ME$(q)$, then
\begin{equation}
\label{Eq.:ratiorate3}
\frac{\Gamma_{\Xi_b \rightarrow D_s^- \Xi_c (1)}}{\Gamma_{\Xi_b \rightarrow \pi^- \Xi_c (1)}}= \frac{p_{D_s^-}(1)\; \overline{\sum}\sum |t|^2(1,D_s^-)}{p_{\pi^-}(1)\; \overline{\sum}\sum |t|^2(1,\pi^-)} =0.686.
\end{equation}
For the semileptonic case, using Eq.\eqref{Eq.:ratiomass}, we have 
\begin{equation}
R= \frac{\Gamma_{\Xi_b \rightarrow \bar{\nu}_l l \ \Xi_c (2790)}}{\Gamma_{\Xi_b \rightarrow \bar{\nu}_l l \ \Xi_c (2815)}} = \frac{\int {\rm d} M_{\rm inv} \; \frac{{\rm d}\Gamma}{{\rm d}M_{\rm inv}}(1)}{\int {\rm d} M_{\rm inv}\; \frac{{\rm d}\Gamma}{{\rm d}M_{\rm inv}}(2)}=0.197.
\end{equation}

We can also estimate the values of the absolute rates by comparing our results with the ones in Refs. \cite{weihong,weisemi} and experimental data from the PDG \cite{pdg}.
For the nonleptonic case we have
\begin{equation}\label{eq:BR_ratio}
  \frac{BR(\Xi_b \to \pi^-\, \Xi^*_c)}{BR(\Lambda_b \to \pi^-\, \Lambda^*_c)}
  = \frac{M_{\Xi^*_c}}{M_{\Xi_b}}\; \frac{M_{\Lambda_b}}{M_{\Lambda^*_c}} \;\frac{q \;\overline{\sum}\sum |t|^2 \Big|_{\Xi_{b}}}{q \;\overline{\sum}\sum |t|^2 \Big|_{\Lambda_b}}\cdot \frac{\Gamma_{\Lambda_b}}{\Gamma_{\Xi_b}},
\end{equation}
where $\overline{\sum}\sum |t|^2 \Big|_{\Lambda_b}$ is given by Eqs.~(41), (42) of Ref.~\cite{weihong}.
Then, from Ref.\cite{pdg} we have the experimental branching ratios of $\Lambda_b \to \pi^- \Lambda_c^*$ and $\frac{\Gamma_{\Lambda_b}}{\Gamma_{\Xi_b}}$, so
 we can obtain the absolute rates:
\begin{eqnarray}\label{eq:BR_Xib}
  BR[\Xi_b \to \pi^- \Xi_c(2790)] &=& (7\pm 4)\times 10^{-6},  \label{eq:BR_Xia1}\\
  BR[\Xi_b \to \pi^- \Xi_c(2815)] &=& (13\pm 7)\times 10^{-6}.  \label{eq:BR_Xia2}
\end{eqnarray}
We can do the same for the semileptonic case where the experimental branching ratios are also in \cite{pdg}
, and comparing with \cite{weisemi}, we obtain:
\begin{eqnarray}\label{eq:BR_PDG}
  BR[\Xi_b \to \bar \nu_l l \Xi_c(2790)] &=& \left( 1.0  ^{+0.6}_{-0.5}  \right) \times 10^{-4}, \label{eq:BR_Xib1} \\
  BR[\Xi_b \to \bar \nu_l l \Xi_c(2815)] &=& \left( 3.3 ^{+1.8}_{-1.6} \right) \times 10^{-4}.\label{eq:BR_Xib2}
\end{eqnarray}

\section{Conclusion}
We have studied the decay of $\Xi_b$ to $\Xi_c^{*}$, through the emission of $\pi^- (Ds^-)$ or $\bar{\nu}_l l$, estimating the ratios of rates and the absolute rates of these reactions. To do that we assumed the resonances were dynamically generated from PB and VB interactions \cite{romanets}, where channels involved were $D \Lambda$, $D \Sigma$, $D^* \Lambda$, $D^* \Sigma$. Here the VB channels were found to be just as important as the PB ones. This confirmation of the importance of VB channels should enhance the necessity of mixing PB and VB to build molecular baryonic states. There are no experiments to confirm our calculations, but our results are within measurable range and can serve as predictions which can also be an incentive to carry new experiments.

\section*{Acknowledgments}
R. P. Pavao wishes to thank the Generalitat Valenciana in the program Santiago Grisolia.
This work is partly supported by the
National Natural Science Foundation of China under Grants No. 11565007, No. 11647309 and No. 11547307.
This work is also partly supported by the Spanish Ministerio
de Economia y Competitividad and European FEDER funds
under the contract numbers  FIS2014-51948-C2-1-P, and FIS2014-51948-C2-2-P, and the Generalitat Valenciana
in the program Prometeo II-2014/068.

\end{document}